\documentclass[12pt,letterpaper]{article}
\usepackage[margin=0.9in]{geometry}
\usepackage[latin1]{inputenc}
\usepackage{amsmath}
\usepackage{amsfonts}
\usepackage{amssymb}
\usepackage{graphicx}
\usepackage{pdfpages}
\author{Alberto Sirlin \\ \\
		{\it Department of Physics, New York University,}\\
	{\it 4 Washington Place, New York, NY 10003 USA}}
\title{Remembering a Great Teacher}

\begin{document}

\maketitle

\abstract{The article is a recollection of the memorable experience of attending a course on Quantum Mechanics given by Feynman in Brasil, as well as several meetings and exchanges Daniele Amati and I had with him over many years, in both the U.S. and Europe. The article also includes a small sample of the problems assigned in the course, a one-page guide, hand-written by Feynman, to study QED on the basis of two of his most important papers, and his reply to a letter of congratulations that the author had sent to him on the occasion of his Nobel Prize Award.}	
	
\vspace*{2.0cm}

	\noindent
	As evidenced by his famous Lectures on Physics, brilliant short books, and video presentations, Richard Feynman was not only one of the greatest physicists of the 20th century, but also a remarkable teacher. This article is a recollection of the memorable experience of attending a course on Quantum Mechanics given by Feynman in Brasil in 1953, as well as several meetings and  exchanges my close friend and classmate Daniele Amati and I had with him over many years, in both the U.S. and Europe. \\ \\
	\noindent
	In order to explain the long and fruitful relation Daniele and I had with Feynman, a brief description of our background is useful. Daniele and I were classmates in high school and university, and completed our doctoral studies at the University of Buenos Aires in 1952. Our research adviser was Richard Gans, a noted classical physicist who was both a theorist and an experimentalist. The subject of our theses was the theory of ferroresonant circuits, an interesting topic in classical physics involving strong non-linearities. As we were quite young at the time (Daniele was 21 years old and I barely 22), the advice of our teachers was to continue our studies abroad to improve our knowledge of Quantum Mechanics and become acquainted with frontier areas such as Quantum Electrodynamics (QED) and Nuclear Physics. Since my hope was to continue my studies in the U.S., I applied to the Institute of International Education, an institution that, among several tasks, served as a liason with american universities. But then, something unexpected happened. One of our teachers, professor Estrella Mathov, was a cosmic-rays physicist and, in her visits to international observatories, had become acquainted with brasilian physicists. Professor Mathov recommended Daniele and me to her brasilian colleagues and, lo and behold, we received offers of one-year fellowships to study at the Centro Brasileiro de Pesquisas Fisicas (Brasilian Center of Physical Research) in Rio de Janeiro. \\ \\
	\noindent
	Daniele and I arrived in Rio in January, 1953. The Centro was a quite remarkable institution. The director, Cesar Lattes, had been a co-discoverer (with Occhialini and Powell) of the pion in cosmic rays and ``O Cesar'' was a household name in Brasil. Among several other people, there were two distinguished brasilian theorists: J.J. Leite Lopes and J. Tiomno, who had been a top student at Princeton and worked with J.A. Wheeler. Leon Rosenfeld was visiting and gave us a very interesting course on classical Statistical Mechanics. Gehrt Moli\`ere, a founder of the theory of multiple scattering, was working at the Centro. David Bohm was teaching in Sao Paulo and visited the Centro to give very interesting talks. There was a steady flow of seminars and short-term visitors, including sometimes very famous people like J.R. Oppenheimer.\\ \\
	\noindent
	Leite Lopes, who was a very lucid and inspiring lecturer, guided our studies. One good day, he gave us the great news: Richard Feynman was planning to visit the Centro again and give a course on Quantum Mechanics. Leite, as we fondly called him, emphasized what a great opportunity this meant for us and the other young students at the Centro.\\ \\
	\noindent
	Finally, the great man arrived: he was about 34 years old, at the peak of his intellectual powers. After all, this was only 4 years after 1949, when he published some of his most important papers in QED and Quantum Field Theory. \\ \\
	\noindent
	Feynman lost no time in spelling out the rules of the course: \\ 
	1) English was forbidden. In particular, Feynman was going to lecture in Portuguese or, more precisely, in Feynmanian Portuguese. \\
	2) It was essential not to miss classes. \\
	3) It was crucial to do the homework assignments. \\
	4) At the end of every class, a student had the right to ask questions about any previous class, provided he had been present at the time.\\ \\
	\noindent
	The number of students attending was quite small. Among them, I remember Sam Mac Dowell, Erasmo Ferreira, Luis Carlos Gomez, an american student, Dobbs, and, of course, Daniele and me. \\ \\
	\noindent
	In the first class, Feynman made a brief survey of what was known and what was not. Leite, who attended that class, noted that Helium superfluidity was not included among the non-understood problems, in contrast with lectures given in previous years. The reply was that significant progress had been recently made. (I think he had in mind his own work on the subject).  \\ \\
	\noindent
	At the beginning, Feynman gave three one-hour lectures a week. Since the course was so special, after a short time we asked him whether he could extend the classes to one hour and a half. Instead, he began to give one-hour lectures five days a week, Mondays through Fridays. Although he was a natural lecturer and could think very fast on his feet, Feynman was not an improviser: his subject material was carefully written in a notebook. In class, Daniele and I took turns: one of us took notes, the other listened. At night, back in the pension where we lived, we improved the notes with the information provided by the listener. Feynman realized what we were doing and told us it was a good idea.\\ \\
	\noindent 
	In the course, Feynman started with the path-integral formulation. After deriving the Schroedinger Equation, the topics and the discussions became closer to more conventional expositions. \\ \\
	\noindent
	The homework problems were interesting and instructive. Since Daniele and I did not master written Portuguese and English was not allowed, we wrote the problems in Spanish and Feynman, who did not have graders, corrected and graded them in his inimitable Portuguese. In particular, he would answer related questions we raised in the process of working out the problems. He liked the way I did the homework and gave me high grades, including an $A^{+}$ and an Otimo (Excellent). He also found my name funny, so one good day he told me: ``Mr. Sirloin, you do your problems very well!'' I did not find out what a sirloin is until I arrived in the U.S. more than one year later, so the pun on my name completely escaped me. \\ \\
	\noindent
	At one time during the course, it was winter-break at the University of Buenos Aires and Daniele and I wanted to spend a week in Argentina to visit our families and friends. Unfortunately, we made the serious error of asking permission to Feynman to skip classes. He was adamant: no, don't go. We went anyway, but brought our notebooks with us. In the first class after our return, we raised our hands to ask a question. Feynman was quite upset: no, you missed the last week of classes. We protested: our question was about a class we attended, namely about the mathematical foundations of the path-integral, and he accepted it. \\ \\
	\noindent
	At the Centro there were two types of seminars: the regular ones, for established physicists, and ``seminarinhos'' or ``small seminars'', for students like us. Daniele and I were invited to give ``seminarinhos'' on the work on ferroresonant circuits we had done in Argentina. To our great surprise, Feynman attended our talks. He asked many questions. We answered perhaps half of them as well as we could; he answered the others himself. But, at the end of our lectures, he told us that he wanted to talk to us. It turns out that we had developed an approximate analytic method, involving harmonics and sub-harmonics, to analyze the solutions of the non-linear differential equations in the resonance region, and Feynman wanted to know to what extent our approach was a good approximation. On the spot, he taught us how to integrate the equations numerically and told us to compare the analytical and numerical solutions. With our talks and the discussion with Feynman, Daniele and I had enough intellectual excitement for one day and, as good Latin Americans, we decided: ``we start tomorrow''. So, some time later, we were chatting in the library. All of a sudden, Feynman materialized in the library. Upon seeing us, he inquired: ``have you started?'' (as if this were the most urgent problem). We mumbled some excuse, while Feynman searched on the shelves for accurate tables to help us in the numerical calculations. (In those days computers were not available to us and numerical calculations were rather painful). When Feynman left the library, Daniele and I told to each other: ``we better start right away, otherwise this guy is going to be very angry!''. In a few days we had a comparison of the analytical and numerical calculations and showed it to Feynman. He seemed to be quite satisfied. From the audience, Feynman was also very active at the regular seminars. During the lectures he used to engage the speakers in intense and sometimes sharp discussions. I remember one occasion in which he went to the blackboard and improvised impromptu explanations and derivations on the subject that were much simpler and clearer than those presented by the lecturer. Interestingly, there were two seminar speakers that could stand their ground before Feynman. One was David Bohm, who gave me the impression of being a very strong and thoughtful physicist. I also believe that Feynman had a lot of respect for him. The other one was M\'{a}rio Schenberg, a big, cigar-smoking Brasilian theorist with a strong mathematical background and a self-confidence that matched Feynman's. \\ \\
	\noindent
	\begin{figure}[!pt]
		\centering
		\includegraphics[width=1\textwidth]{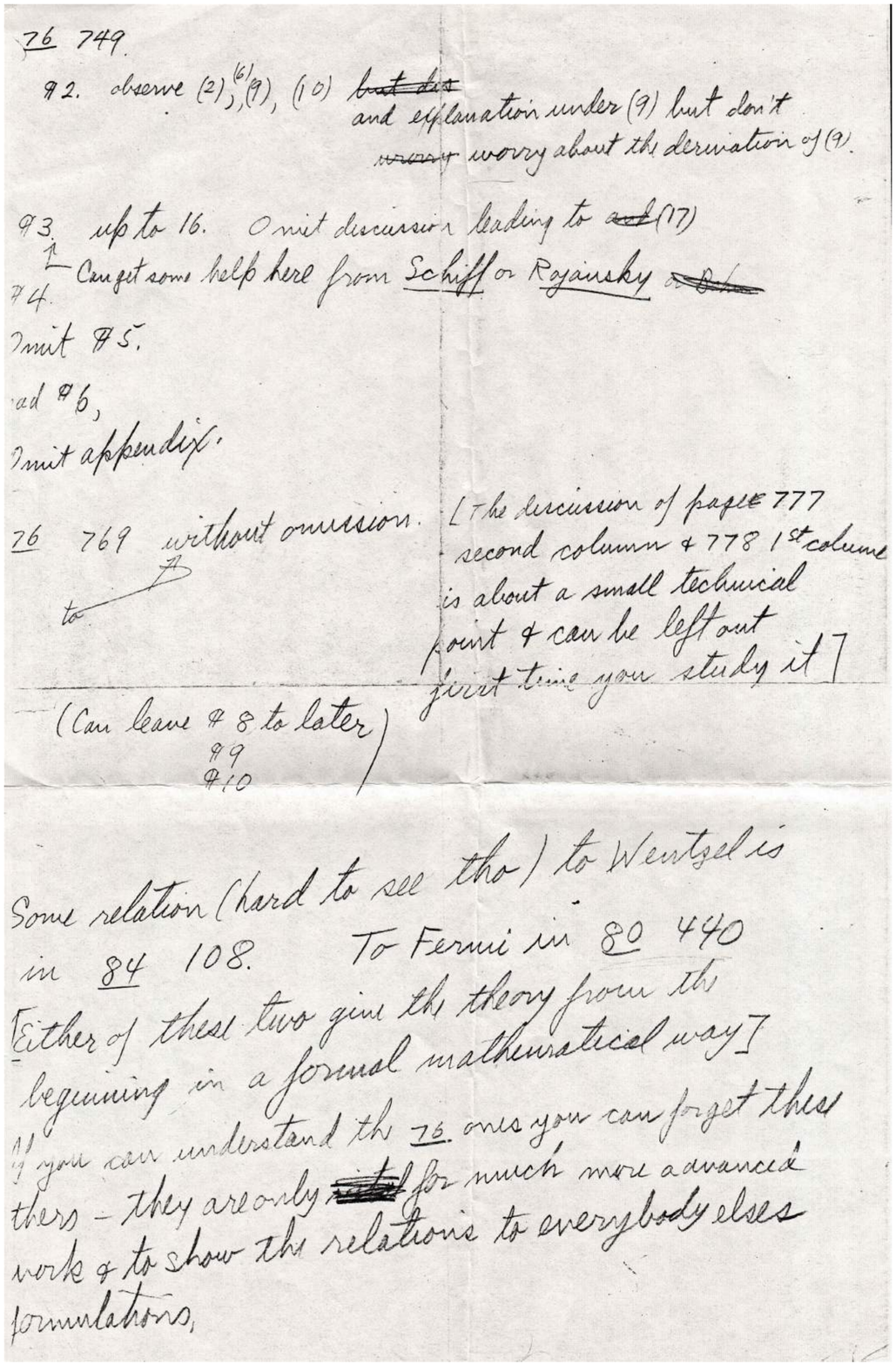}
		\caption{One-page guide, hand-written by Feynman. to study QED on the basis of two of his great 1949 papers. \label{fig:guide}}
	\end{figure}
	Towards the end of the course, Daniele and I approached Feynman. We told him that we were very grateful for the wonderful course he had given us, but then we posed the question: ``what shall we do about QED?''. Feynman left immediately for his office and wrote a one-page guide to study QED on the basis of two of his greatest papers: ``The Theory of Positrons'', Physical Review 76, 749 (1949) and ``Space-Time Approach to Quantum Electrodynamics'', Physical Review 76, 769 (1949). (See Figure~\ref{fig:guide}.) The guide is very detailed. For example, it indicates the equations, derivations and Sections that can be omitted in the first reading, as well as references to other formulations.  He also told us that he would mail to the Centro copies of his lectures on QED. After his stay in Brasil, Feynman travelled to Japan (I believe that he participated in a conference there). A short time later, several copies of his QED lectures reached the Centro. Their content is very similar to that of his book ``Quantum Electrodynamics'' (The Benjamin/Cummings Publishing Company, Inc., Reading,Massachusetts, 1961). I picked up one copy and began to study it systematically.\\ \\
	\noindent
	At the end of our one-year fellowships, Daniele and I returned to Buenos Aires. There I learned that my application to the Institute of International Education had been successful: UCLA informed me that I had been accepted as a graduate student and offered a teaching assistanship. I wrote to Feynman with the good news and he answered back with some advice: i) don't be afraid to ask questions, no matter how elementary they may sound and ii) you may learn as much by discussing physics with other graduate students as with professors. Then he added: when you settle down in Los Angeles, give me a call.\\ \\
	\noindent
	I arrived in Los Angeles in September, 1954. This was my first trip outside South America. Everything seemed so different that, at first, I had the feeling that I had landed in a different planet! Fortunately, after a short while, I was well settled and phoned Feynman, who invited me to visit him at Caltech. This turned out to be an exciting and memorable visit. Feynman started by describing his recent research, which was mostly on the superfluidity of Helium. He then asked me about my own work. I explained that I had been studying the energy-angle distribution of the radiation emitted by a relativistic charged particle traversing a thin target, a combination of multiple-scattering and radiation theories, a topic that had been suggested to me by Moli\`ere in Brasil, and had also been discussed by Leonard Schiff. Feynman thought that this work might be of interest to some experimentalists and took me to visit a laboratory and talk briefly with some of the physicists there. Back in his office, the conversation continued and at one point Feynman phoned his wife and invited me for dinner at his home. After dinner, it was getting late and Feynman was kind enough to take his car and drive me back to my apartment at the edge of the UCLA campus, all the while continuing our very interesting conversation. \\ \\
	\noindent
	At UCLA, I attended my first formal course on Quantum Field Theory given by Robert Finkelstein (a remarkable theorist who even now, at age 99, continues to publish sophisticated papers on particle physics), as well as a course on Nuclear Physics. On the research side, I collaborated with another graduate student, Ralph Behrends, and professor Finkelstein in calculating the radiative corrections of $O(\alpha)$ to muon decay in the framework of the four-fermion Fermi theory of weak interactions. This work, published in 1956, was one of the first to apply the powerful relativistic methods developed by Feynman, Schwinger, Tomonaga, Dyson and others in QED to evaluate the radiative corrections to weak interactions. An important issue is that, in evaluating the radiative corrections to muon decay, one must generalize the Feynman integrals in QED to the case when the virtual charged fermion changes its mass at the weak interaction vertex. I remember that, as a graduate student, I was very excited at the thought that, in some sense, we were generalizing one of Feynman's great inventions! \\ \\
	\noindent
	Meanwhile, in 1954 Daniele returned to Italy, his native country, and in the summer he attended a one-month course given by Fermi. He wrote me an interesting letter comparing the two as teachers. He put it roughly as follows: you ask a question to Feynman and, in many cases, he invents on the spot a clever and very original argument to answer the question; you understand the argument but at the same time you have the strong feeling that in no way you would have thought of it. You ask a question to Fermi and his answer is so simple that you have the strong feeling that you should have thought of it yourself!  \\ \\
	\noindent
	\begin{figure}[t]
		\centering
		\includegraphics[width=0.48\textwidth]{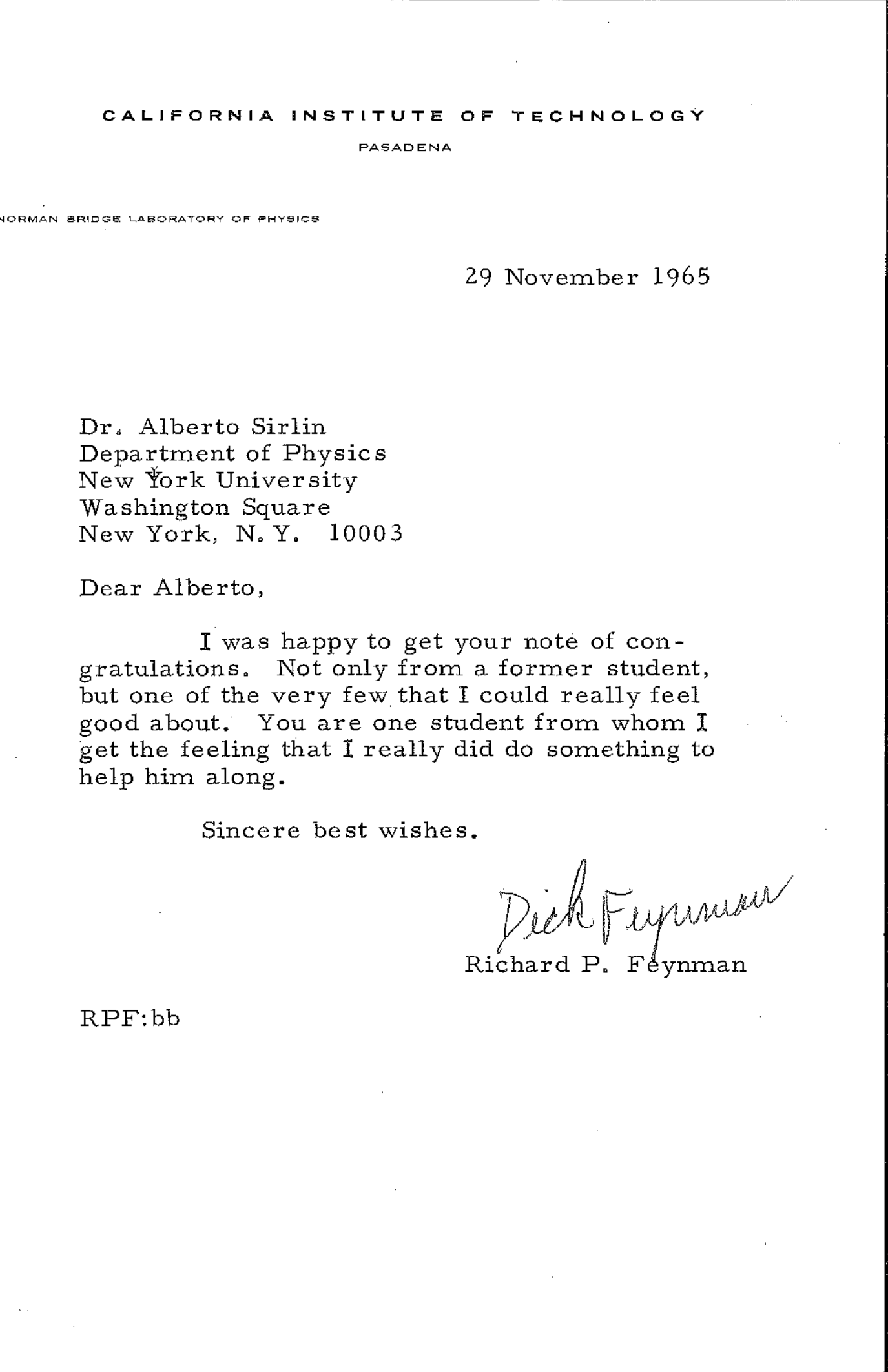}
		\caption{Reply to a letter of congratulations that the author had sent to Feynman on the occasion of his Nobel Prize Award. \label{fig:letter}}
	\end{figure}
	A short time after CERN was founded, Daniele joined its theory division. Feynman visited CERN a number of times and in such occasions they got together. In his letters, Daniele wrote that Feynman repeatedly expressed the opinion that we were the students that had really learned from him. At first, I found this strange because it was natural to imagine that at Caltech he most likely had some excellent students. After a while, I thought that I understood the basis of his opinion. It turns out that, when we were students at the University of Buenos Aires, except for a few lectures, we had not attended a full, systematic course on Quantum Mechanics. A good part of our knowledge was self-taught using well-known textbooks such as Schiff's. In particular, we were not aware of the path-integral formulation of Quantum Mechanics. In contrast, several of his other students had already taken a full course on Quantum Mechanics, and it was not clear to Feynman what was his own impact on their knowledge. In the case of Daniele and me, he felt that his course in Brasil filled this gap in our education in Quantum Mechanics, and therefore we ``were the students that had really learned from him''.  My late colleague Engelbert Schucking, who was a great scholar of physics and read everything, told me that there is a biography of Feynman, ``To the Beat of a Different Drum'', by the historian J. Mehra, where he is asked about his graduate students. He replied that there were three students on whom he felt he had a very positive influence: Daniele and me in Brasil, and Sam Berman, one of his students in Caltech. When Feynman was awarded the Nobel Prize, I wrote a short letter asking him to accept the warmest congratulations from one of his former students. He wrote back a short, very nice letter (see Figure~\ref{fig:letter}), that includes the following interesting remark: ``Not only from a former student, but one of the very few that I could really feel good about. You are one student from whom I get the feeling that I really did do something to help him along''. Aside from his memorable course on Quantum Mechanics, another very important contribution Feynman made to my education as a young physicist, were the QED lectures he mailed to the Brasilian Center at his return to Caltech. Since the theory of precision electroweak physics became one of the main focus of my research, those notes turned out to be most useful to me.\\ \\
	\noindent
	A very nice quality of Feynman was his readiness to acknowledge the contributions of others. In 1960 he was summarizing the results obtained by Sam Berman (his student at Caltech) and by Toichiro Kinoshita and me (at Cornell and Columbia) on the radiative corrections of $O(\alpha)$ to muon and beta decays in the Vector - Axialvector (V - A) theory of charged-current weak processes that had been proposed by Feynman and Gell-Mann and Sudarshan and Marshak in the late fifties. At first hand, the results were disconcerting: the radiative corrections to muon decay were finite while those for beta decay were logarithmically divergent (so they were evaluated with a cutoff chosen by intuition). A distinguished physicist in the audience asked what was the reason for this conundrum. It turns out that I had developed a simple argument, based on Fierz transformations and a comparison with QED, that explained the riddle. As Feynman hesitated, Kinoshita told me to explain my argument. While I was leaving the room at the end of his lecture, Feynman told me: ``Very good, Alberto; some time ago I had the same argument but I had forgotten it''. On a couple of occasions I met Feynman during summer visits to the Aspen Center for Physics. In one of these visits, he tried to practice his Spanish with me. He explained that Mexican physicists had invited him and, knowing that in Brasil he had lectured in Portuguese, they had asked him to give his Mexican lectures in Spanish!. I also tried to convince him to lecture in New York. Unfortunately, he could not make it at that time, but a few years later he did lecture at Columbia University on the parton model. The large auditorium was completely filled. Apparently, a large part of the audience were people from other departments in the university who were attracted by his great fame as a lecturer.\\ \\
	\noindent
	In 1974, in another visit to Aspen, I participated in a workshop organized by Murray Gell-Mann on various topics of interest at the time. It turns out that I had recently re-examined the problem of the radiative corrections to beta decay in the framework of the Standard Model (SM). Since this is a renormalizable theory, I argued with myself that I should get a finite answer. I first considered a simplified version of the SM with integer-charged quarks, neglecting the effect of the strong interactions, and found that the logarithmic divergence of the radiative corrections evaluated in the Fermi theory of weak interactions was replaced by a finite, large logarithm of the form $\ln(M_Z/m_p)$ where $M_Z$ and $m_p$, are the Z-boson and proton masses. At the Aspen workshop, I was invited to talk on this work. Unfortunately, Feynman was not able to attend my lecture. However, the next day I was driving my old rented car when I saw Feynman walking in the same direction and offered him a lift. He picked up a cup of coffee at a nearby store and joined me in the car. He then told me: ``Murray told me that you solved the old problem of the divergence of the radiative corrections to beta decay''. In later work (published in 1978) I extended the result to the real SM with fractionally charged quarks, taking into account the effect of the strong interactions using a current algebra formulation. A very interesting feature is that the radiative corrections obtained in this way are not only finite but also large which, in combination with the experimental data, is necessary to ensure the unitarity of the first row of the Cabibbo-Kobayashi-Maskawa quark mixing matrix.\\ \\
	\noindent
	The last time I met Feynman was in the 1983 Shelter Island II conference, organized in commemoration of the historic 1947 Shelter Island conference, in which several great physicists discussed their fundamental contributions to QED. In the 1983 conference, also attended by several scientific luminaries, Feynman gave a lecture on the strong interactions. Towards the end of the conference, I approached Feynman to greet him and express my thanks for the new lecture. He was wearing rather thick eyeglasses and, at first, it seemed that he did not recognize me. When I reminded him that I was one of his former students in Brasil, he answered: ``Of course, I know who you are. I always say that those two (referring to Daniele and me) were the students that really learned from me''.\\ \\
	\noindent
	\section*{Acknowledgments}
	The writing of this article was supported in part by the National Science Foundation under Grant No. PHY-1316452. The author is greatly indebted to Daniele Amati for several communications and suggestions, and to Andrea Ferroglia for very useful discussions. 
	
\appendix 

\section*{Appendix}

This Appendix includes a small sample	of the problems assigned by Feynman in the course on Quantum Mechanics described in the article. 

\begin{enumerate}
	\item[1)] If the nucleus is a uniform sphere of radius $a = \frac{e^2}{2 m c^2} A^{1/3}$, what is the modification of the energy levels of H in first order for the $1s,2s,2p$ states? Also, if possible, derive a general expression for the higher states.
	\item[2)] There is a positive charge $Ze$ and a system of electronic charges with density $\rho(\mathbf{r})$. Show that:
	\begin{displaymath}
	v\left(\mathbf{q}\right) = \frac{4 \pi \hbar^2 e^2}{\mathbf{q}^2} \left(Z-\int e^{\frac{i}{\hbar} \mathbf{q} \cdot \mathbf{r}} \rho \left(\mathbf{r} \right) d^3 \mathbf{r} \right) 
	\end{displaymath} 
	\item[3)] Demonstrate:
	\begin{displaymath}
	K = F\left(t_1,t_2\right) e^{\frac{i}{\hbar} S_{\text{cl}}}
	\end{displaymath}
	when 
	\begin{displaymath}
	L = \alpha(t) \dot{x}^2 + \beta(t) x \dot{x} +\gamma(t) x^2 +A(t) \dot{x} +B(t) x +C(t)
	\end{displaymath}
	\item[4)] Calculate $K$ for the free particle in 3 dimensions.
	\item[5)] Show for the harmonic oscillator $L = \frac{m}{2} \left( \dot{x}^2 - \omega^2 x^2 \right)$,
	that
	\begin{displaymath}
	K \left(x_2,t_2;x_1,t_1 \right) = 
\sqrt{\frac{m \omega}{2 \pi \hbar i \sin(\omega T)}} e^{\frac{i}{\hbar} S_{\text{cl}} (x_2,t_2;x_1,t_1)}\, ,
	\end{displaymath}
\end{enumerate}
where $T = t_2 - t_1$.

The boldface letters stand for three-dimensional vectors.
In problems 3,4,5, $K$ is the space-time propagator evaluated in the path integral formulation and, in the second problem, $v(\mathbf{q})$ is the Fourier transform of the electrostatic potential energy $V(\mathbf{s})$ in the interaction between the charge distribution and an external electron. These problems and their solutions are shown below.
In these hand-written notes double lined letters stand for three-dimensional vectors. 
Since English was not allowed in the course, Daniele and I wrote the text of the problems in Spanish while Feynman graded them in Portuguese.

\newpage


\section*{Supplementary material}

1) Si el núcleo es una esfera uniforme de radio $a = \frac{1}{2}\frac{e^2}{mc^2}A^{1/3}$, ¿cuál es la modificación de los niveles del H en primer orden para los estados 1s, 2p, 2s? e indicar, si se puede, una expresión general para los estados siguientes.

Para $r > a$, el potencial no queda alterado por el tamaño finito del núcleo:
$V(r) = -\frac{Ze^2}{r}$ para $r > a$

Para $r < a$, el campo eléctrico es el que corresponde a una esfera de densidad uniforme de radio $r$. La carga de esta esfera es $\frac{Ze r^3}{a^3}$ y la energía potencial que ella determina para $r < a$ será $+\frac{Ze^2 r^2}{2a^3} - \frac{3Ze^2}{2a}$

Luego $V(r) = \begin{cases} -\frac{Ze^2}{r} & \text{para } r > a \\ -\frac{Ze^2}{r} + \underbrace{\frac{Ze^2}{r} + \frac{Ze^2 r^2}{2a^3} - \frac{3Ze^2}{2a}}_{U} & \text{para } r < a \end{cases}$

Como queremos estudiar la desviación con respecto al caso Coul. puro, podemos definir nuestro potencial perturbador $U$ de la siguiente manera:

$U(r) = \begin{cases} 0 & \text{para } r > a \\ Ze^2\left[\frac{1}{r} + \frac{r^2}{2a^3} - \frac{3}{2a}\right] & \end{cases}$

Luego, en primer orden, la variación de energía para el estado $n, \ell$ será:

$\Delta E = U_{n\ell;n\ell} = \int_{-\infty}^{+\infty} \psi^*_{n\ell m} U(r) \psi_{n\ell m} d^3x = \int_0^a (R_{n\ell}(r))^2 U(r) r^2 dr$

debemos tomar $R_{n\ell}$ normalizada según $\int_0^\infty (R_{n\ell})^2 r^2 dr = 1$

En tales condiciones:

$R_{n\ell} = -\left\{\left(\frac{2Z}{na_0}\right)^3 \frac{(n-\ell-1)!}{2n((n+\ell)!)^3}\right\}^{1/2} e^{-\rho/2} \rho^\ell L_{n+\ell}^{2\ell+1}(\rho)$ donde $\boxed{\rho = \frac{2Zr}{na_0}}$

$\Delta E = U_{n\ell;n\ell} = Ze^2 \left(\frac{2Z}{na_0}\right)^3 \frac{(n-\ell-1)!}{2n((n+\ell)!)^3} \int_0^a e^{-\rho} \rho^{2\ell}\left(L_{n+\ell}^{2\ell+1}(\rho)\right)^2 \left(r + \frac{r^4}{2a^3} - \frac{3r^2}{2a}\right) dr$

$\boxed{= \frac{Z^2 e^2}{n^2 a_0} \frac{(n-\ell-1)!}{((n+\ell)!)^3} \int_0^\alpha e^{-\rho}\rho^{2\ell+1}\left(L_{n+\ell}^{2\ell+1}(\rho)\right)^2 \left(1 - \frac{3\rho}{2\alpha} + \frac{\rho^3}{2\alpha^3}\right) d\rho = \Delta E} \quad \boxed{\alpha = \frac{2Za}{na_0}}$

Naturalmente, dado $n$ y $\ell$ esta integral puede calcularse exactamente por integraciones parciales. Sin embargo, podemos hacer algunas aproximaciones para obtener fácilmente una fórmula general. Para ello basta observar que $\alpha$ es muy pequeño pues $a = 1.4 \times 10^{-13} A^{1/3}$ cm y $a_0 \simeq 0.53 \times 10^{-8}$ cm. Cuanto más liviano sea el núcleo más pequeño será $\alpha$. Por lo tanto en el integrando podemos reemplazar $e^{-\rho}$ por 1 y $L_{n+\ell}^{2\ell+1}(\rho)$ por el término de orden cero de su desarrollo en potencias de $\rho$, es decir por $L_{n+\ell}^{2\ell+1}(0)$.

Calcularemos ahora cuánto vale $L_n^m(0)$

$$L_n^m(\rho) = \frac{d^m}{d\rho^m} L_n(\rho) = \frac{d^m}{d\rho^m}\left(\frac{d}{d\rho}-1\right)^n \rho^n = (-1)^n \frac{d^m}{d\rho^m}\left(1-\frac{d}{d\rho}\right)^n \rho^n$$

$$\left(1-\frac{d}{d\rho}\right)^n = 1 - \binom{n}{1}\frac{d}{d\rho} + \binom{n}{2}\frac{d^2}{d\rho^2} + \cdots = \sum_{j=0}^{n}\binom{n}{j}\frac{d^j}{d\rho^j}(-1)^j$$

en que $\binom{n}{j} = \frac{n!}{(n-j)!j!}$ es decir $\binom{n}{j}$ representa un "número combinatorio".

Luego $L_n^m(\rho) = \sum_{j=0}^{n-m} \frac{n!}{(n-j)!j!}(-1)^{j+n}\frac{d^{j+m}}{d\rho^{j+m}}\rho^n =$

$\sum_{j=0}^{n-m} \frac{n!}{(n-j)!j!}(-1)^{j+n}\frac{n!}{(n-j-m)!}\rho^{(n-j-m)}$ ; Si llamo $k = n-j-m$:

$$\boxed{L_n^m(\rho) = \sum_{k=0}^{n-m}(-1)^{k+m}\frac{(n!)^2 \rho^k}{(k+m)!(n-k-m)!k!}}$$

$$L_n^m(0) = (-1)^m\frac{(n!)^2}{m!(n-m)!} \qquad \therefore \boxed{L_{n+\ell}^{2\ell+1}(0) = (-1)^{2\ell+1}\frac{(n+\ell)!^2}{(2\ell+1)!(n-\ell-1)!}}$$

Llamando $E_n = -\frac{Z^2 e^2}{2a_0 n^2}$ (energía del caso Coul. puro):

$$\Delta E = -2E_n \frac{(n+\ell)!}{((2\ell+1)!)^2(n-\ell-1)!}\int_0^\alpha \rho^{2\ell+1}\left(1-\frac{3\rho}{2\alpha}+\frac{\rho^3}{2\alpha^3}\right)d\rho$$

$-2E_n\frac{(n+\ell)!}{((2\ell+1)!)^2(n-\ell-1)!}\cdot \frac{3}{(2\ell+2)(2\ell+3)(2\ell+5)}\alpha^{2\ell+2} = -\frac{6 E_n(n+\ell)!\alpha^{2(\ell+1)}}{(2\ell+5)(n-\ell-1)!(2\ell+1)!(2\ell+3)!}$

$$\boxed{(\Delta E)_{n\ell} = -\frac{6}{(2\ell+5)}\frac{E_n(n+\ell)!}{(n-\ell-1)!(2\ell+1)!(2\ell+3)!}\left(\frac{2Z}{n}\frac{a}{a_0}\right)^{2(\ell+1)}}$$

Caso 1s para H $(n=1, \ell=0, Z=1)$

$(\Delta E)_{10} = -\frac{4}{5}\left(\frac{a}{a_0}\right)^2 E_1 \quad\therefore \frac{(\Delta E)_{10}}{E_1} = -\frac{4}{5}\left(\frac{1.4\times 10^{-13}}{0.53\times 10^{-8}}\right)^2 \sim -5.6\times 10^{-10}$ ✓

$\boxed{\frac{\Delta E_{10}}{E_1} \sim -5.6\times 10^{-10}}$ ; $E_1 = -$ Rydberg $= -13.6$ volts $\therefore \boxed{(\Delta E)_{10}\sim 7.61\times 10^{-9}\text{ volts}}$

Para el estado 2s $(n=2; \ell=0)$:

$\frac{(\Delta E)_{20}}{E_2} = -\frac{2}{5}\left(\frac{a}{a_0}\right)^2 \sim \boxed{-2.8\times 10^{-10} = \frac{(\Delta E)_{20}}{E_2}}$ $E_2 = -\frac{\text{Ryd}}{4} = -3.4$ volts $\therefore$

$\boxed{(\Delta E)_{20} \sim 9.5\times 10^{-10}\text{ volts}}$ ✓

Para el estado $2p$ $(n=2, \ell=1)$:

$$\frac{(\Delta E)_{21}}{E_2} = -\frac{1}{140}\left(\frac{a}{a_0}\right)^4 \sim \boxed{-3.5 \times 10^{-21} = \frac{(\Delta E)_{21}}{E_2}}$$

$$\boxed{(\Delta E)_{21} \sim 11.9 \times 10^{-21} \text{ evolts}}$$

En general, observamos que la influencia del radio finito del núcleo es pequeñísima siendo mayor en el estado fundamental $1s$. Esta influencia disminuye un poco al pasar al estado $2s$ y disminuye notablemente al considerar el estado $2p$. Esto está de acuerdo con nuestras ideas generales pues sabemos que cuando $r \to 0$ la función de onda se comporta como $r^\ell$ y por tanto podemos predecir que la influencia del radio finito del núcleo debe ser necesariamente mayor para los estados $s$ que para los siguientes. Esto se debe a que cuando aumenta el momento angular, el electrón no puede acercarse mucho al origen (es decir, hay poca probabilidad de que se acerque).

Podemos considerar todos los estados $s$ ($\ell=0$): ✓

$$\boxed{\frac{(\Delta E)_{n0}}{E_n} = -\frac{4}{5}\frac{Z^2}{n}\left(\frac{a}{a_0}\right)^2} \checkmark$$

fórmula que muestra que para $\ell=0$, la influencia relativa del radio finito del núcleo sobre los niveles de energía decrece como $1/n$ al aumentar $n$. La modificación absoluta de los niveles variará como $\frac{1}{n^3}$. Las influencias del radio finito del núcleo aumentan, si consideramos núcleos pesados, como $Z^{2(\ell+1)}$.

Observación.: Hemos usado una teoría de perturbaciones para niveles no degenerados, en ciertos estados degenerados, del átomo de H. A primera vista esto parecería incorrecto pues de acuerdo con aquella teoría teníamos en la primera aproximación: $a_m(E_0-E_n) = V_{0n}$; por tanto, si $E_0 = E_n$ (estados degenerados) debe ser $V_{0n}=0$. Si esta condición no se cumple falla el método usado. Pero en nuestro caso esta condición se cumple, es decir, $V_{0n}=0$. Más aún, es claro que siempre que $V(r)$ sean esféricos simétricos $V_{0n}=0$. En el problema 3 trataremos el efecto Stark y veremos que allí no podemos usar la teoría para niveles no degenerados pues en ese caso el potencial $V(r)$ no será radial. Otimo! ←

1) Se tiene una carga positiva $Ze$ y alrededor un sistema de cargas electrónicas con densidad $\rho(\vec{r})$. Demostrar:

$$v(\vec{q}) = \frac{4\pi \hbar^2 e^2}{q^2}\left(Z - \int e^{\frac{i}{\hbar}\vec{q}\cdot\vec{r}} \rho(\vec{r}) d^3\vec{r}\right)$$

La carga positiva $Ze$ actúa como un potencial Coulomb y según vimos en clase su contribución a $v(\vec{q})$ es $\frac{4\pi\hbar^2 e^2 Z}{q^2}$.

~~Las contribuciones a $v(\vec{q})$ de la carga positiva y de la nube electrónica son aditivas ya que $v(\vec{q})$ depende linealmente del potencial $V$.~~
Estudiemos ahora la contribución de la nube electrónica.

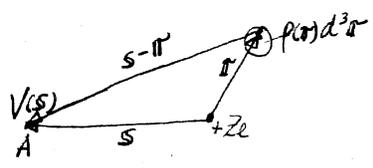

Al punto $A$ donde se produce el scattering, es decir donde se calcula $V$, le asigno el vector $\vec{s}$ por definición.

Por definición:
$$v(\vec{q})_{el.} = \int e^{\frac{i}{\hbar}\vec{q}\cdot\vec{s}} V(\vec{s}) d^3\vec{s}$$

Calculemos $V(\vec{s})$. El elemento de volumen $d^3\vec{r}$ contribuye en $\vec{s}$ un potencial $-\frac{e^2 \rho(\vec{r}) d^3\vec{r}}{|\vec{s}-\vec{r}|}$. Luego la nube electrónica origina en $\vec{s}$ un potencial

$$V(\vec{s})_{el.} = -e^2 \int \frac{\rho(\vec{r})}{|\vec{s}-\vec{r}|} d^3\vec{r}$$

$$v(\vec{q})_{el.} = -e^2 \iint \frac{e^{\frac{i}{\hbar}\vec{q}\cdot\vec{s}} \rho(\vec{r})}{|\vec{s}-\vec{r}|} d^3\vec{r}\, d^3\vec{s}$$

Calculemos la integral respecto a $\vec{s}$:

$$\int \frac{e^{\frac{i}{\hbar}\vec{q}\cdot\vec{s}}}{|\vec{s}-\vec{r}|} d^3\vec{s} = \int \frac{e^{\frac{i}{\hbar}\vec{q}\cdot(\vec{s}-\vec{r}+\vec{r})}}{|\vec{s}-\vec{r}|} d^3(\vec{s}-\vec{r}) =$$

$$e^{\frac{i}{\hbar}\vec{q}\cdot\vec{r}} \int \frac{e^{\frac{i}{\hbar}\vec{q}\cdot\vec{v}}}{|\vec{v}|} d^3\vec{v} \qquad \text{donde } \vec{v} = \vec{s}-\vec{r}$$

Elegimos coordenadas polares esféricas de modo que $\vec{q}$ sea el eje polar:

$$= e^{\frac{i}{\hbar}\vec{q}\cdot\vec{r}} \iint \frac{e^{\frac{i}{\hbar}qv\cos\theta}}{v} v^2 d(\cos\theta)\, dv = 2\pi e^{\frac{i}{\hbar}\vec{q}\cdot\vec{r}} \int_0^\infty dv\, v \left(\frac{e^{\frac{i}{\hbar}qv} - e^{-\frac{i}{\hbar}qv}}{\frac{i}{\hbar}qv}\right) =$$

$$= \frac{2\pi e^{\frac{i}{\hbar}\vec{q}\cdot\vec{r}}}{\frac{i}{\hbar}q} \cdot 2i \int_0^\infty \sen\left(\frac{qv}{\hbar}\right) dv = \frac{4\pi e^{\frac{i}{\hbar}\vec{q}\cdot\vec{r}}}{q^2} \hbar^2 \int_0^\infty \sen\left(\frac{qv}{\hbar}\right) d\left(\frac{qv}{\hbar}\right) =$$

$$= \frac{4\pi \hbar^2}{q^2} e^{\frac{i}{\hbar}\vec{q}\cdot\vec{r}} \cos\left(\frac{qv}{\hbar}\right)\Bigg|_{\frac{qv}{\hbar}=\infty}^{\frac{qv}{\hbar}=0} = \frac{4\pi \hbar^2}{q^2} e^{\frac{i}{\hbar}\vec{q}\cdot\vec{r}}$$

Luego
$$\boxed{v(\vec{q})_{el} = -\frac{4\pi \hbar^2 e^2}{q^2} \int e^{\frac{i}{\hbar}\vec{q}\cdot\vec{r}} \rho(\vec{r}) d^3\vec{r}} \qquad \boxed{v(\vec{q})_{tt} = \frac{4\pi \hbar^2 e^2}{q^2}\left(Z - \int e^{\frac{i}{\hbar}\vec{q}\cdot\vec{r}} \rho(\vec{r}) d^3\vec{r}\right)}$$

**Problema**: Demostrar $K = F(t_1, t_2) e^{\frac{i}{\hbar} S_{cl}}$ cuando

$$L = \alpha(t)\dot{x}^2 + \beta(t) x \dot{x} + \delta(t) x^2 + A(t)\dot{x} + B(t) x + C(t)$$

Trayect. clásica: $\dfrac{d}{dt}\left(\dfrac{\partial L}{\partial \dot{x}}\right) - \dfrac{\partial L}{\partial x} = 0 \quad \therefore$

$$2\alpha \ddot{\bar{x}} + 2\dot{\alpha}\dot{\bar{x}} + 2\dot{\bar{x}}\alpha + \bar{x}\dot{\beta} + \dot{A} - B = 0$$

$$K = \int_{t_1} e^{\frac{i}{\hbar}\int_{t_1}^{t_2} L\, dt}\, \mathcal{D}_{x(t)} \qquad \text{Haciendo } x = \bar{x} + y$$

$$S = \int_{t_1}^{t_2} L\, dt = \int_{t_1}^{t_2}(\alpha \dot{x}^2 + \beta x \dot{x} + \delta x^2 + A\dot{x} + B x + C)\, dt =$$

$$= \int_{t_1}^{t_2}\left[\alpha(\dot{\bar{x}}+\dot{y})^2 + \beta(\bar{x}+y)(\dot{\bar{x}}+\dot{y}) + \delta(\bar{x}+y)^2 + A(\dot{\bar{x}}+\dot{y}) + B(\bar{x}+y) + C\right] dt =$$

$$= \int_{t_1}^{t_2}[\alpha\dot{\bar{x}}^2 + 2\alpha\dot{\bar{x}}\dot{y} + \alpha\dot{y}^2 + \beta\bar{x}\dot{\bar{x}} + \beta\bar{x}\dot{y} + \beta y \dot{\bar{x}} + \beta y \dot{y} + \delta\bar{x}^2 + \delta y^2 + 2\delta\bar{x} y + A\dot{\bar{x}} + A\dot{y} + B\bar{x} + B y + C]\, dt$$

Siendo $\int_{t_1}^{t_2} 2\alpha\dot{\bar{x}}\dot{y}\, dt = [2\alpha\dot{\bar{x}} y]_{t_1}^{t_2} - 2\int y\ddot{\bar{x}}\alpha\, dt - 2\int y\dot{\bar{x}}\dot{\alpha}\, dt$

$$\int_{t_1}^{t_2}\beta\bar{x}\dot{y} = [\beta\bar{x} y]_{t_1}^{t_2} - \int (y\beta\dot{\bar{x}} + y\bar{x}\dot{\beta})\, dt$$

$$\int A\dot{y}\, dt = [A y]_{t_1}^{t_2} - \int y \dot{A}\, dt \qquad \text{se obtiene}$$

$$S = \int_{t_1}^{t_2}[\alpha\dot{\bar{x}}^2 - 2 y\ddot{\bar{x}}\alpha - 2 y\dot{\bar{x}}\dot{\alpha} + \alpha\dot{y}^2 + \beta\bar{x}\dot{\bar{x}} - y\beta\dot{\bar{x}} - y\bar{x}\dot{\beta} + \beta y\dot{\bar{x}} + \beta y\dot{y} + \delta\bar{x}^2 + \delta y^2 + 2\delta\bar{x} y + A\dot{\bar{x}} - \dot{A} y +$$
$$+ B\bar{x} + B y + C]\, dt =$$

$$= \int_{t_1}^{t_2}[-y(2\alpha\ddot{\bar{x}} + 2\dot{\alpha}\dot{\bar{x}} + \bar{x}\dot{\beta} - 2\delta\bar{x} + \dot{A} - B) + (\alpha\dot{\bar{x}}^2 + \beta\bar{x}\dot{\bar{x}} + \delta\bar{x}^2 + A\dot{\bar{x}} + B\bar{x} + C) + (\alpha\dot{y}^2 + \beta y\dot{y} + \delta y^2)]\, dt$$

La ecuación del movimiento clásico anula el primer paréntesis; el segundo paréntesis es el lagrangeano clásico $\therefore$

$$S = S_{cl} + \int_{t_1}^{t_2}(\alpha\dot{y}^2 + \beta y\dot{y} + \delta y^2)\, dt \qquad \text{y por ello}$$

$$K_{(1\to 2)} = \int e^{\frac{i}{\hbar} S_{cl}} \cdot e^{\frac{i}{\hbar}\int_{t_1}^{t_2}(\alpha\dot{y}^2 + \beta y\dot{y} + \delta y^2)\, dt}\, \mathcal{D}_{y(t)} = e^{\frac{i}{\hbar} S_{cl}}\int e^{\frac{i}{\hbar}\int_{t_1}^{t_2}(\quad)\, dt}\, \mathcal{D}_{y(t)}$$
$$\scriptsize y=0 \text{ en } t_1,\ y=0 \text{ en } t_2$$

siendo la integral función de $t_1$ y $t_2$ solamente

$$K_{(1,2)} = F(t_1, t_2) e^{\frac{i}{\hbar} S_{cl}} \qquad \text{ó} \qquad K_{(1,2)} = \varphi(t_2 - t_1) e^{\frac{i}{\hbar} S_{cl}}$$

La forma de la función $\varphi(t_2-t_1)$ se determina, como ha sido hecho para la part. libre y para el oscilador armónico, por el principio de combinación de las amplitudes.

**Problema.** Calcular $K$ para la partícula libre en 3 dimensiones.

$K = \int e^{\frac{i}{\hbar}S} \mathcal{D}[\vec{r}(t)]$ ; $L = \frac{m}{2}(\dot{x}^2 + \dot{y}^2 + \dot{z}^2)$

Hacemos: $\begin{cases} x = \bar{x} + X \\ y = \bar{y} + Y \\ z = \bar{z} + Z \end{cases}$   Eq. del movim. clásico: $\begin{cases} \ddot{\bar{x}} = 0 \\ \ddot{\bar{y}} = 0 \\ \ddot{\bar{z}} = 0 \end{cases}$   $\begin{aligned} \dot{\bar{x}} &= \frac{x_2 - x_1}{t_2 - t_1} = cte \\ \dot{\bar{y}} &= \frac{y_2 - y_1}{t_2 - t_1} \\ \dot{\bar{z}} &= \frac{z_2 - z_1}{t_2 - t_1} \end{aligned}$

$$S = \int_{t_1}^{t_2} L\, dt = \frac{m}{2} \int_{t_1}^{t_2} \{(\dot{\bar{x}} + \dot{X})^2 + (\dot{\bar{y}} + \dot{Y})^2 + (\dot{\bar{z}} + \dot{Z})^2\}\, dt = \frac{m}{2} \int_{t_1}^{t_2} [\dot{\bar{x}}^2 + \dot{\bar{y}}^2 + \dot{\bar{z}}^2]\, dt$$

$$+ \frac{m}{2} \cdot 2 \int_{t_1}^{t_2} [\dot{\bar{x}}\dot{X} + \dot{\bar{y}}\dot{Y} + \dot{\bar{z}}\dot{Z}]\, dt + \frac{m}{2} \int_{t_1}^{t_2} (\dot{X}^2 + \dot{Y}^2 + \dot{Z}^2)\, dt$$

Pero $\int \dot{\bar{x}} \dot{X}\, dt = \int \dot{\bar{y}} \dot{Y}\, dt = \int \dot{\bar{z}} \dot{Z}\, dt = 0$ por lo ya visto.

Luego
$$S = \frac{m}{2} \int_{t_1}^{t_2} (\dot{\bar{x}}^2 + \dot{\bar{y}}^2 + \dot{\bar{z}}^2)\, dt + \frac{m}{2} \int_{t_1}^{t_2} (\dot{X}^2 + \dot{Y}^2 + \dot{Z}^2)\, dt$$

$$K(2,1) = \iiint e^{\frac{i}{\hbar} S(2,1)} \mathcal{D}[X] \mathcal{D}[Y] \mathcal{D}[Z]$$

$X = 0$ en $t_1, t_2$ ; $Y = 0$ en $t_1, t_2$ ; $Z = 0$ en $t_1, t_2$

$$K(2,1) = e^{\frac{m}{2} \frac{i}{\hbar} \int_{t_1}^{t_2} (\dot{\bar{x}}^2 + \dot{\bar{y}}^2 + \dot{\bar{z}}^2)\, dt} \iiint e^{\frac{m}{2} \frac{i}{\hbar} \int_{t_1}^{t_2} (\dot{X}^2 + \dot{Y}^2 + \dot{Z}^2)\, dt} \mathcal{D}[X] \mathcal{D}[Y] \mathcal{D}[Z]$$

↳ función sólo de $t_1$ y $t_2$.

$$\boxed{K(2,1) = \varphi(t_2 - t_1)\, e^{\frac{i}{\hbar} S_{cl}}}\quad \text{con}$$

$$S_{cl} = \frac{m}{2} \frac{\{(x_2-x_1)^2 + (y_2-y_1)^2 + (z_2-z_1)^2\}}{t_2 - t_1} = \frac{m}{2} \frac{(\vec{r}_2 - \vec{r}_1)^2}{t_2 - t_1}$$

Calculemos $\varphi(t_2 - t_1)$. Usando el principio de combinación de amplitudes:

$$K(2,1) = \iiint_{-\infty}^{+\infty} dx_3\, dy_3\, dz_3\, K(3,1) K(2,3)$$ obtenemos como hicimos para una dimensión:

$$\boxed{\varphi(t_2 - t_1) = \left(\frac{m}{2\pi i \hbar (t_2 - t_1)}\right)^{3/2}}$$

Entonces:
$$\boxed{K(2,1) = \left(\frac{m}{2\pi i \hbar (t_2 - t_1)}\right)^{3/2} e^{\frac{i}{\hbar} \frac{m}{2} \frac{(\vec{r}_2 - \vec{r}_1)^2}{t_2 - t_1}}}$$

Problema 1)

Mostrar para el oscilador armónico $L = \frac{m}{2}(\dot{x}^2 - \omega^2 x^2)$, que
$$K(x_2, t_2; x_1, t_1) = \sqrt{\frac{m\omega}{2\pi \hbar i \operatorname{sen}\omega T}}\, e^{\frac{i}{\hbar} S_{cl}(x_2, t_2, x_1, t_1)}$$

Hagamos $x(t) = \bar{x}(t) + y(t)$; $\frac{d}{dt}\left(\frac{\partial L}{\partial \dot{x}}\right) - \frac{\partial L}{\partial x} = 0 \longrightarrow \ddot{\bar{x}} = -\omega^2 \bar{x}$

$$K = \int e^{\frac{i}{\hbar}\int_{t_1}^{t_2} L\, dt}\, \mathcal{D}[x(t)] = \int e^{\frac{i}{\hbar}\int_{t_1}^{t_2}\frac{m}{2}(\dot{x}^2 - \omega^2 x^2)\, dt}\, \mathcal{D}[x(t)]$$

(todas tray. de 1 a 2) (tray.)

$$I = \int_{t_1}^{t_2}(\dot{x}^2 - \omega^2 x^2)\, dt = \int_{t_1}^{t_2}\left[(\dot{\bar{x}} + \dot{y})^2 - \omega^2[\bar{x} + y]^2\right] dt =$$

$$\int_{t_1}^{t_2}\left[(\dot{\bar{x}})^2 + 2\dot{\bar{x}}\dot{y} + \dot{y}^2 - \omega^2\bar{x}^2 - \omega^2 y^2 - 2\omega^2 \bar{x} y\right] dt$$

Pero $\int_{t_1}^{t_2}\dot{\bar{x}}\dot{y}\, dt = [y\dot{\bar{x}}]_{t_1}^{t_2} - \int y\ddot{\bar{x}}\, dt$
(por partes)   $\longrightarrow 0$ porque $y=0$ en los límites.

Luego:
$$I = \int_{t_1}^{t_2}[\dot{\bar{x}}^2 - 2y\ddot{\bar{x}} + \dot{y}^2 - \omega^2\bar{x}^2 - \omega^2 y^2 - 2\omega^2\bar{x}y]\, dt =$$
$$\int_{t_1}^{t_2}[\dot{\bar{x}}^2 - 2y\underbrace{(\ddot{\bar{x}} + \omega^2\bar{x})}_{0\,(tray.\,clásica)} + \dot{y}^2 - \omega^2\bar{x}^2 - \omega^2 y^2]\, dt = \int_{t_1}^{t_2}[(\dot{\bar{x}}^2 - \omega^2\bar{x}^2) + (\dot{y}^2 - \omega^2 y^2)]\, dt$$

Luego $S = \int_{t_1}^{t_2} L\, dt = \frac{m}{2}\int_{t_1}^{t_2}(\dot{x}^2 - \omega^2 x^2)\, dt = \frac{m}{2}\int_{t_1}^{t_2}(\dot{\bar{x}}^2 - \omega^2\bar{x}^2) + \frac{m}{2}\int_{t_1}^{t_2}(\dot{y}^2 - \omega^2 y^2)\, dt =$
$S_{cl} + \frac{m}{2}\int(\dot{y}^2 - \omega^2 y^2)\, dt$

Queda entonces: $K = \int e^{\frac{i}{\hbar}S}\, \mathcal{D}[x(t)] = e^{\frac{i}{\hbar}S_{cl}} \int e^{\frac{i}{\hbar}\frac{m}{2}\int_{t_1}^{t_2}(\dot{y}^2 - \omega^2 y^2)\, dt}\, \mathcal{D}[y(t)]$

(todas tray. entre 1 y 2)  (tray. $y=0$ en $t_1, t_2$; función sólo de $t_1$ y $t_2$)

Luego $\boxed{K = F(t_1, t_2)\, e^{\frac{i}{\hbar}S_{cl}} = F(t_1,t_2)\, e^{\frac{im}{\hbar 2}\int_{t_1}^{t_2}(\dot{\bar{x}}^2 - \omega^2\bar{x}^2)\, dt}}$

donde $F(t_1, t_2)$ es una función de $t_2 - t_1$.
Calculando $S_{cl}$ para el oscilador armónico teníamos:
$$S_{cl} = \frac{m\omega}{2\,\operatorname{sen}\omega(t_2 - t_1)}\left[(x_1^2 + x_2^2)\cos\omega(t_2 - t_1) - 2x_1 x_2\right]$$

Determinemos ahora la función $F(t_1,t_2) = \varphi(t_2 - t_1)$. Por el principio de combinación de amplitudes:

$$K(2,1) = \int_{-\infty}^{+\infty} K(3,1)\, K(2,3)\, dx_3$$

$$\psi(t_2-t_1) e^{\frac{i}{\hbar}\frac{m\omega}{2\sen\omega(t_2-t_1)}[(x_1^2+x_2^2)\cos\omega(t_2-t_1)-2x_1 x_2]}$$

$$= \int_{-\infty}^{+\infty} \psi(t_3-t_1)\psi(t_2-t_3) e^{\frac{i}{\hbar}\frac{m\omega}{2\sen\omega(t_3-t_1)}[(x_1^2+x_3^2)\cos\omega(t_3-t_1)-2x_1 x_3]} e^{\frac{i}{\hbar}\frac{m\omega}{2\sen\omega(t_2-t_3)}[(x_3^2+x_2^2)\cos\omega(t_2-t_3)-2x_3 x_2]} dx_3 =$$

$$= \psi(t_3-t_1)\psi(t_2-t_3) e^{\frac{i}{\hbar}\frac{m\omega}{2}\left[\frac{x_1^2\cos\omega(t_3-t_1)}{\sen\omega(t_3-t_1)} + \frac{x_2^2\cos\omega(t_2-t_3)}{\sen\omega(t_2-t_3)}\right]}$$

$$\times \int_{-\infty}^{+\infty} e^{\frac{i}{\hbar}\frac{m\omega}{2}\left(\frac{\cos\omega(t_3-t_1)}{\sen\omega(t_3-t_1)} + \frac{\cos\omega(t_2-t_3)}{\sen\omega(t_2-t_3)}\right)x_3^2 - \frac{i}{\hbar}\frac{m\omega}{2}\left[\frac{2x_2}{\sen\omega(t_2-t_3)} + \frac{2x_1}{\sen\omega(t_3-t_1)}\right]x_3} dx_3$$

Por la integral I.1:
$$= \psi(t_3-t_1)\psi(t_2-t_3) e^{\frac{i}{\hbar}\frac{m\omega}{2}\left[\frac{x_1^2\cos\omega(t_3-t_1)}{\sen\omega(t_3-t_1)} + \frac{x_2^2\cos\omega(t_2-t_3)}{\sen\omega(t_2-t_3)}\right]} \sqrt{\frac{\pi i 2\hbar \sen\omega(t_3-t_1)\sen\omega(t_2-t_3)}{m\omega \sen\omega(t_2-t_1)}} \times$$

$$\times \exp\left[-\frac{i m\omega}{2\hbar} \frac{[x_2\sen\omega(t_3-t_1)+x_1\sen\omega(t_2-t_3)]^2}{\sen\omega(t_2-t_3)\sen\omega(t_3-t_1)\sen\omega(t_2-t_1)}\right]$$

Las exponenciales se cancelan. Queda:

$$\psi(t_2-t_1) = \psi(t_3-t_1)\psi(t_2-t_3) \sqrt{\frac{\pi i 2\hbar \sen\omega(t_3-t_1)\sen\omega(t_2-t_3)}{m\omega \sen\omega(t_2-t_1)}}$$

Haciendo la sustitución:

$$\varphi(t_j-t_i) = \sqrt{\frac{m\omega}{\pi i 2\hbar \sen\omega(t_j-t_i)}} \, \psi(t_j-t_i) \text{ obtenemos:}$$

$$\varphi(t_2-t_1) = \varphi(t_3-t_1)\varphi(t_2-t_3) \quad \therefore \quad \varphi \text{ es la exponencial:}$$

$$\varphi(t_2-t_1) = e^{k(t_2-t_1)}, \ldots \varphi(T) = e^{kT}. \text{ Eligiendo } k=0: \varphi(T)=1$$

$$\boxed{\psi(T) = \sqrt{\frac{m\omega}{2\pi\hbar i}} \frac{1}{\sqrt{\sen\omega T}}} \quad \boxed{K(2,1) = \sqrt{\frac{m\omega}{2\pi\hbar i \sen\omega(t_2-t_1)}} e^{\frac{i}{\hbar}S_{cl}}}$$

$$\boxed{K(2,1) = \sqrt{\frac{m\omega}{2\pi\hbar i \sen\omega(t_2-t_1)}} e^{\frac{i}{\hbar}\frac{m\omega}{2\sen\omega(t_2-t_1)}[(x_1^2+x_2^2)\cos\omega T - 2x_1 x_2]}}$$


	
\end{document}